\documentclass[twocolumn]{revtex4}

\usepackage{subfigure}
\usepackage{epsfig}

\usepackage{graphicx}
\usepackage{dcolumn}
\usepackage{bm}

\begin{document}

\title { Electromagnetically induced transparency  in two channels of a $\Lambda$ system }
\author{S. M. Iftiquar}
\affiliation{ Department of Physics, Indian Institute of Science, Bangalore 560012, India.}

\begin{abstract}
 We show two channel electromagnetically induced transparency (EIT) spectra at various optical power. The two channels are coupling field and probe field absorption. It is shown that EIT width and intensity increases linearly with pump power. A density matrix calculation is used to determine nature of EIT spectra induced by optical field. It has been observed that EIT occurs in cases when coupling laser is at fixed frequency while probe laser frequency is scanned, when coupling laser is frequency scanned and probe is at fixed frequency.  
\end{abstract}

\maketitle

\section{Introduction}

Recently a great deal of interest has been drawn on electromagnetically induced transparency because of its potential application in photon storage [1-4], manipulation and coherently transfer of information from photon to atom [5-6], all these are important steps in quantum information processing. In a single photon process atom strongly absorbs resonant photon but by absorption it is transferred from stable ground state to relatively unstable excited state from where is decays rapidly. Thus in a single photon absorption the information carried by photon is rapidly dissipated. In 1997 Harris [7] proposed a scheme in which when two laser fields are applied to atomic system the atom, under certain condition, transfer its coherence between its two transition pathways in such a way that the photonic information remains trapped within the atomic system for a longer duration. In [8,9] it was shown that photonic group velocity is indeed slowed down appreciably during the EIT phase.

This EIT phenomena has been studied extensively in the recent times from theoretical as well as experimental point of view. Theoretical analysis showed that the electromagnetically induced transparency is very similar to coherent population trapping and several proposals also came up like light deflection [10]  etc, that uses coherent control of photon dynamics in presence of atomic vapor. In the experimental part several attempts have been made to improve the characteristic features  of the EIT spectra like line narrowing [11], slowing down light speed through the medium [12], controlled absorption and transparency of atomic vapor system [13] etc.

It has long been considered that first and third order susceptibility $\chi^{(1)},\ \chi^{(3)}$ can also have a role in EIT [14-16] but as magnitude of $\chi^{(3)}$ is several order less than that of $\chi^{(1)}$ so role of linear susceptibility will be more prominant than nonlinear one.

It has been popularly known that coupling laser creates dressed states that destructively interferes in probe absorption at a particular probe frequency, creating EIT window. Here we theoretically and experimentally show that EIT can also happen when coupling laser frequency ($\omega_{c}$) scans and probe frequency ($\omega_{r}$) is kept fixed.

In the following we derive expression of probe absorption in linear susceptibility regime. Then we demonstrate experimental results and finally show how the theoretical EIT profile matches with the experimentally observed spectra.

  \begin{figure}[h]
\includegraphics[angle=0,width=7cm]{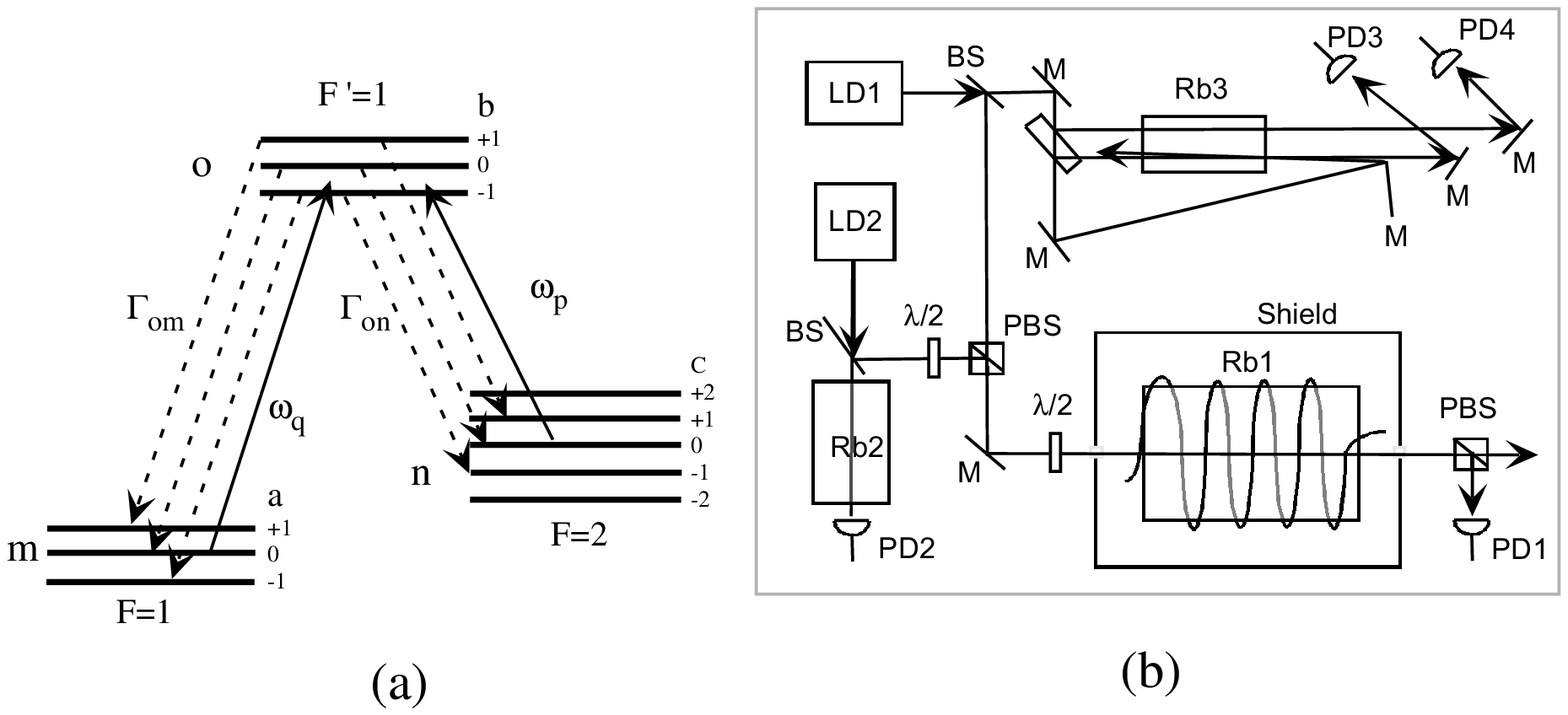}
  \caption{(a) Energy level diagram of $^{87}Rb$ ground states $F=1,\ 2$  and excited state $F^{\prime}=1$ denoted as $| m \rangle,\ |n \rangle$ and $| o \rangle$, along with their Zeeman sublevels $a,\ c,\ b$ respectively. (b) Experimental setup for EIT. LD1 pump and LD2 probe laser, BS beam splitter, PBS polarizing cube beam splitter, $\lambda /2$ half wave plate, M mirror, PD1 EIT detecting photodiode, PD2, PD3, PD4 are for saturated absorption detection, shield is magnetic shield to Rb1 rubidium cell that is used as EIT cell, Rb2 and Rb3 are rubidium cells for saturated absorption spectroscopy.}
  \label{fig1}
  \end{figure}

\section{Theory}

We consider  a $\Lambda$-type bare atomic state that take part in dynamic processes through $^{87}Rb\ (^{85}Rb)$ $D_{2}$ transitions where the two hyperfine (HF) ground states $^{87}Rb, F=1,2\ (^{85}Rb\ F=2,3)$ and one excited HF state $F^{\prime}=1\ (F^{\prime}=2)$ are considered along with its possible Zeeman sublevels. We start the theoretical formulation with $^{87}Rb$ whose energy level structure is shown in figure 1a. We denote $|m\rangle \equiv  F=1 ,\ |n\rangle \equiv  F=2 ,\ |o\rangle \equiv  F^{\prime}=1 $ respectively. EIT is observed when coupling laser drives atom from $ F=2   \rightarrow  F^{\prime}=1 $ in presence of a weaker probe at a frequency $ F=1  \leftrightarrow  F^{\prime}=1 $. We look into the atomic dynamics through interaction picture, where probability amplitudes of atomic states vary slowly with time in rotating wave approximation. We use  $\hat{\rho}$ as atomic density operator, a $3 \times 3$ square matrix with component $\rho_{xy}=| x \rangle \langle y |$
where $|x\rangle,\ |y\rangle  \rightarrow |m\rangle,\ |n\rangle,\ |o \rangle$. Electric dipole moment vector of atom $\vec{\mu}$ has operator form $\hat{\mu}$ with matrix element $\mu_{xy}=-e\langle x | \hat{r} |y\rangle $, the value of which depends on Clebsch Gordan coefficient, $\hat{r}$ is position operator of electron of charge $e$, diagonal components of $\mu=0$. 

In a semiclassical approximation total atomic Hamiltonian $\hat{H}$ is sum of free Hamiltonian $\hat{H}_{0}$ and interaction Hamiltonian $\hat{H}_{I}$, or $\hat{H}=\hat{H}_{0} +\hat{H}_{I}$, [8] where


\begin{eqnarray}
\lefteqn { H_{0} = \hbar \omega_{x}| m \rangle \langle m | + \hbar \omega_{y} |n \rangle \langle n | + \hbar \omega_{z} |o \rangle \langle o |   } \\ \nonumber
& & H_{I} = -\frac{\hbar}{2}(\Omega e^{-i (\phi_{2} +\omega_{p} t}| o \rangle \langle n |) + {\rm H. c.}
\end{eqnarray}

where  $\hbar \Omega =-\vec{\mu}_{no}. \vec{E}(t)$, and $\Omega_{p} e^{-i \phi_{2}}$ is complex Rabi frequency of pump beam, $\omega_{p}$ frequency of pump laser, $\phi_{2}$ its phase. Generally subscript $p$ is used for pump beam and subscript $q$ for probe. Density matrix equation can be written as


\begin{eqnarray}
\dot{\rho}_{xy}= -\frac{i}{\hbar}[\hat{H}, \hat{\rho}]_{xy}  - (\gamma_{xy}+ f)\rho_{yx}  \ \\
 \dot{\rho}_{yy}= -\frac{i}{\hbar}[\hat{H}, \hat{\rho}]_{yy}  +\sum_{E_{x}>E_{y}}\Gamma_{xy}\rho_{xx} - \sum_{E_{x}<E_{y}}\Gamma_{yx}\rho_{yy} 
\end{eqnarray}

$\hbar \omega_{x}$ energy of atomic state $|x \rangle$, $\Gamma_{xy}$ is decay rate of population from $|x\rangle$ to $|y\rangle$, $\gamma_{xy}$ is  decay rate of coherence $\rho_{xy}$ and $\gamma_{yx}=\gamma_{xy}=\frac{1}{2}(\Gamma_{x}+\Gamma_{y})$ where $\Gamma_{x}$ is total population decay rate from $|x\rangle$.  $f$ power broadening or incoherence introduced by laser field to atomic coherence that results in modified $\gamma_{xy} \rightarrow (\gamma_{xy} + f)$,  $f$ is a fraction of Rabi frequency. We find solution of above equations (2,3) by perturbative expansion and find solution for $\rho_{xy}^{(1)}$.

For monochromatic pump field the electric field is same to its Fourier component; $|\vec{E}(t)| = |\vec{E}(\omega_{p})| =E$, say. So $\rho_{xy}^{(1)}$ can be written as


\begin{eqnarray}
\rho^{(1)}_{xy}= \hbar ^{-1} (\rho^{(0)}_{xx}-\rho^{(0)}_{yy})\frac{\vec{\mu}_{xy}.\vec{E}(\omega_{p}) e^{- i \omega_{p}t}}{(\omega_{xy}-\omega_{p}) - i (\gamma_{xy}+f )}  
\end{eqnarray}

 \begin{figure}[h]
 \includegraphics[angle=0,width=5cm]{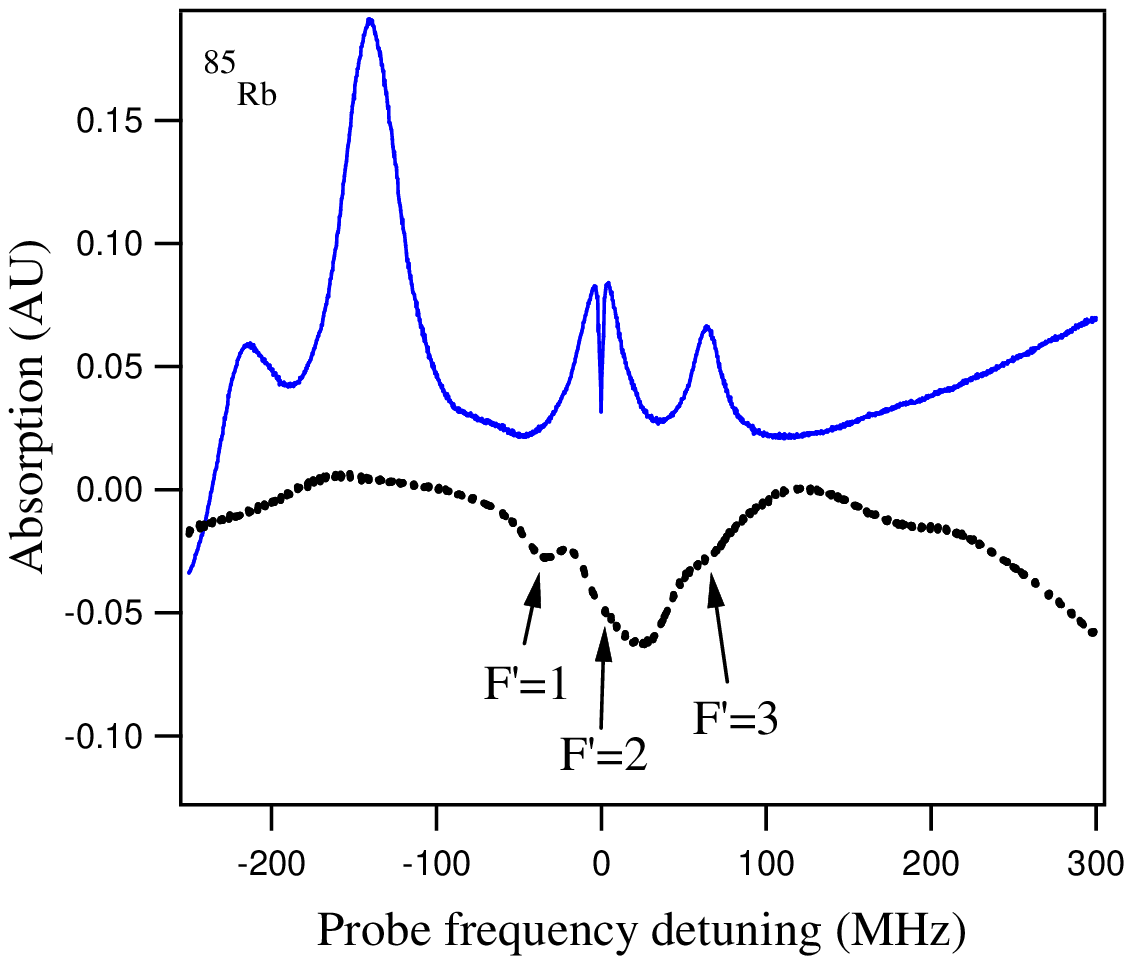}
  \caption{$^{85}Rb$ EIT spectra when when coupling laser connects $F=3 \rightarrow F^{\prime}=2$, and probe laser scans over $F=2 \rightarrow F^{\prime}=1,2,3$. EIT is visible at $F=2 \rightarrow F^{\prime}=2$ transition. The dotted lower curve showssaturated absorption spectra (from photodiode PD2) while upper curve is the EIT spectra (obtained by photodiode PD1). }
 \label{fig2}
  \end{figure}

 \begin{figure}[h]
 \includegraphics[angle=0,width=5cm]{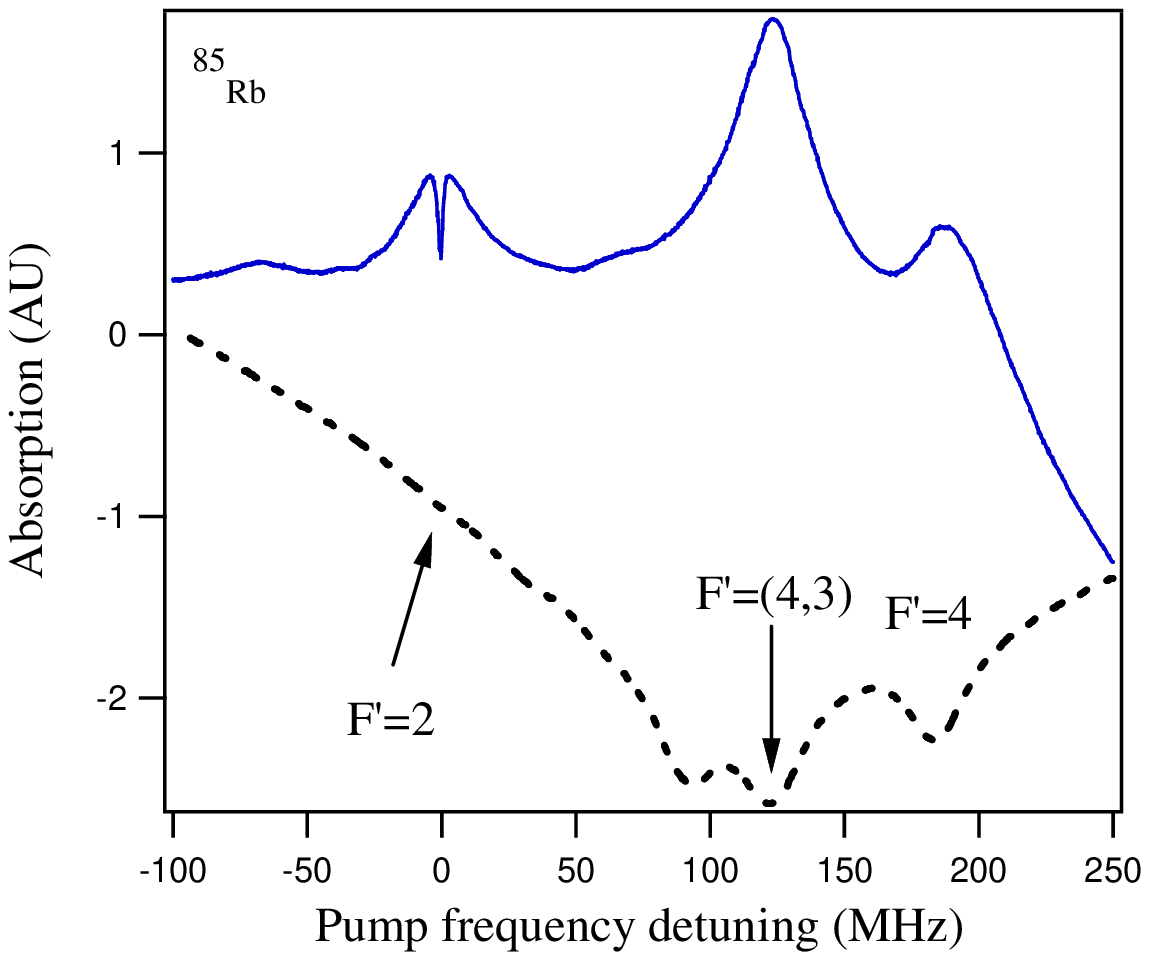}
  \caption{$^{85}Rb$ EIT spectra when when coupling laser connects $F=2 \rightarrow F^{\prime}=2$, and probe laser scans over $F=3 \rightarrow F^{\prime}=2,3,4$. EIT is visible at $F=3 \rightarrow F^{\prime}=2$ transition. The dotted lower curve showssaturated absorption spectra (from photodiode PD2) while upper curve is the EIT spectra (obtained by photodiode PD1).  }
 \label{fig3}
  \end{figure}

 \begin{figure}[h]
 \includegraphics[angle=0,width=5cm]{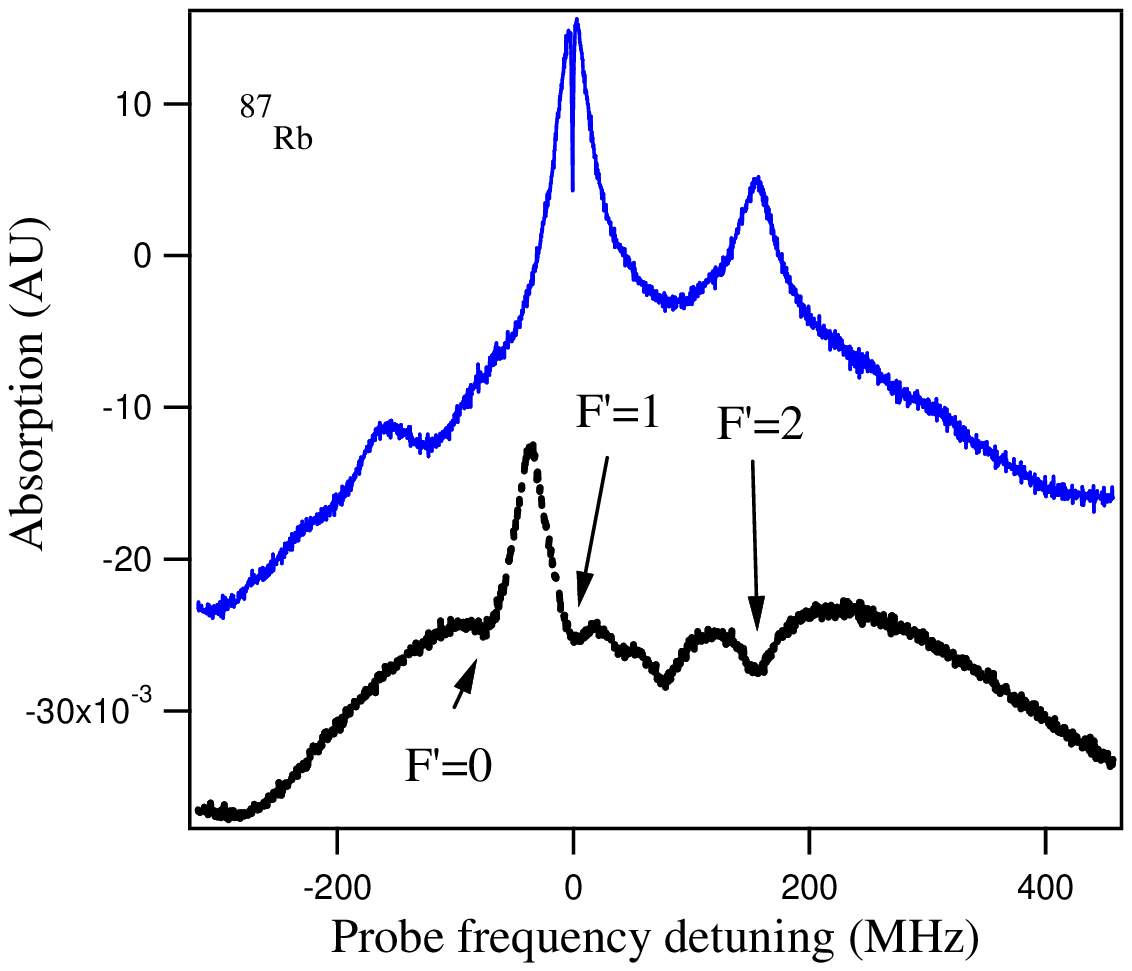}
  \caption{$^{87}Rb$ EIT spectra when when coupling laser connects $F=2 \rightarrow F^{\prime}=1$, and probe laser scans over $F=1 \rightarrow F^{\prime}=0,1,2$. EIT is visible at $F=1 \rightarrow F^{\prime}=1$ transition. The dotted lower curve shows saturated absorption spectra (from photodiode PD2) while upper curve is the EIT spectra (obtained by photodiode PD1).}
 \label{fig4}
  \end{figure}

 \begin{figure}[h]
 \includegraphics[angle=0,width=5cm]{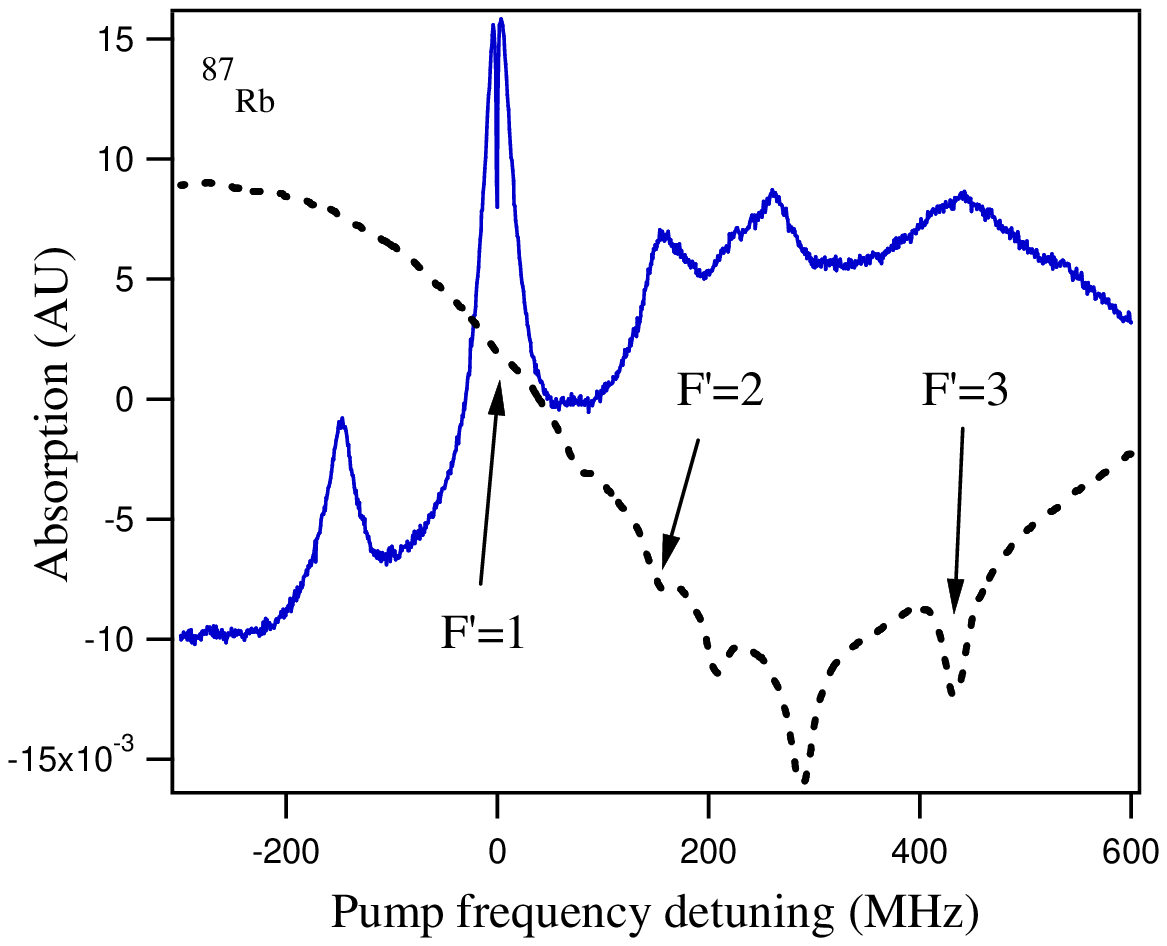}
  \caption{ $^{87}Rb$ EIT spectra when when coupling laser connects $F=1 \rightarrow F^{\prime}=1$, and probe laser scans over $F=2 \rightarrow F^{\prime}=1,2,3$. EIT is visible at $F=2 \rightarrow F^{\prime}=1$ transition. The dotted lower curve shows saturated absorption spectra (from photodiode PD2) while upper curve is the EIT spectra (obtained by photodiode PD1).}
 \label{fig5}
  \end{figure}

 \begin{figure}[h]
 \includegraphics[angle=0,width=7cm]{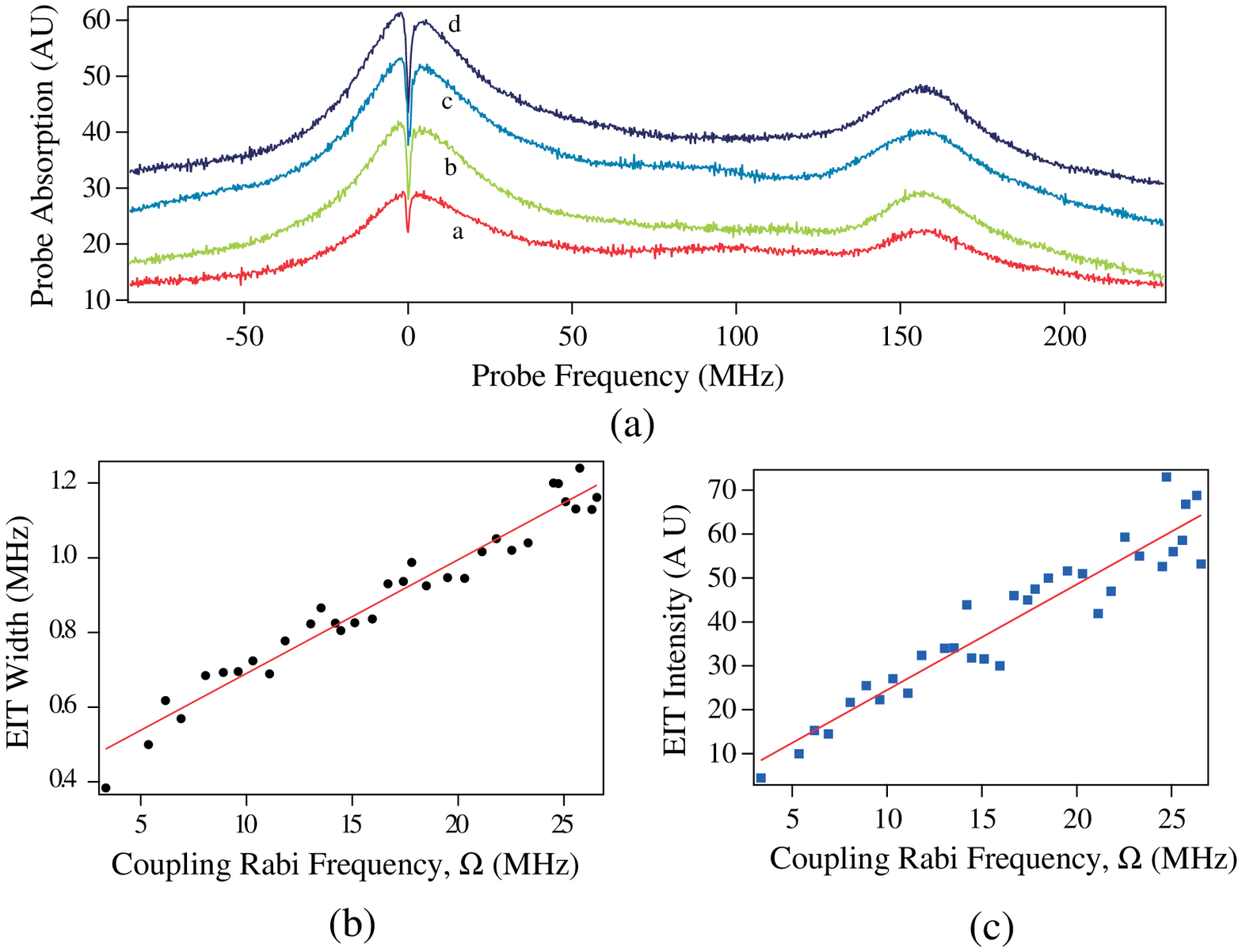}
  \caption{(a) EIT spectra at $|m\rangle \rightarrow |o \rangle$ transition around zero frequency scale, while optically pumped absorption around 157 MHz. curves $a, b, c, d$ are for power $18,\ 46,\ 61,\ 76 \ \mu$Watt. (b) Lorentzian width of EIT dip at various pump power, the line is a theoretical fit. (c) Peak depth of EIT signal at various pump power, the line is a theoretical fit.}
 \label{fig4}
  \end{figure}

Using $\rho^{(1)}$ we determine first order polarization of atomic system the imaginary component of which is 


\begin{eqnarray}
{\rm Im}P^{(1)}_{ij}(\omega_{p})=\frac{N}{\hbar}\sum_{cb}(\rho^{(0)}_{cc}-\rho^{(0)}_{bb})\frac{E \mu_{cb}^{i}\mu_{bc}^{j}}{(\omega_{cb}-\omega_{p})-i(\gamma_{bc}+ f)} \\ \nonumber
{\rm Im}P^{(1)}_{ij}(\omega_{q})=\frac{N}{\hbar}\sum_{ab}(\rho^{(0)}_{aa}-\rho^{(0)}_{bb})\frac{E_{q} \mu_{ab}^{i}\mu_{ba}^{j}}{(\omega_{ab}-\omega_{p})-i(\gamma_{ba}+ f)}
\end{eqnarray}

where $N$ is atomic number density in vapor. We assume $\mu_{cb}^{i}, \mu_{bc}^{j}$ as real and $E \mu_{cb}^{i}=\hbar \Omega_{p}$. The presence of coupling laser results in optical pumping and increased probe absorption, in the linear approximation which can be mathematically expressed as


\begin{eqnarray}
\alpha(\omega_{q})={\rm Im}\left (P_{ij}^{(1)}(\omega_{q})\right ) \left [1+ G \times {\rm Im}\left (P_{ij}^{(1)}(\omega_{p})\right )\right ]
\end{eqnarray}

where $G\leq 1$, is a factor that determines optical pumping. Among all the possible transitions, it is to be noted that $ F^{\prime}=1,m^{\prime}_{z}=0  \rightarrow  F=1, m_{z}=0  $ is dipole forbidden so optical pumping by spontaneous decay from excited state to this level is negligible. Furthermore, probe absorption from $ F=1, m_{z}=0   \rightarrow  F^{\prime}=1,m^{\prime}_{z}=0  $ is also dipole forbidden. So the corresponding coherence decay constant $\gamma_{om}$ is very small; $(\gamma_{om})_{m_{z}\neq 0}\sim 10^{4}(\gamma_{om})_{m_{z}= 0}$, as compared to that of other Zeeman sublevels.

 For $D_{2}$ spectra, Zeeman transition from upper ground state  $^{87}Rb, F=2,m_{F}=0 $ to excited state $ F^{\prime}=1, m_{F}=0 $ is possible but  $ F=1,m_{F}=0  \leftrightarrow  F^{\prime}=1,m_{F}=0 $ transition is dipole forbidden this state does not decay to $F=1,m_{F}=0$, thus $\rho_{aa,m_{F}=0}\sim 0$, where $m_{F}$ is Zeeman quantum number and from now on $a,\ c,\ b $ stands for dummy indices representing Zeeman levels of $|m \rangle,\ |n \rangle,\ |o \rangle$. So, the effect on total probe absorption can be expressed as a convolution of individual absorption from the Zeeman levels
 

\begin{eqnarray}
\lefteqn { \alpha(\omega_{q})= \left ( -A_{q}\frac{(\gamma_{ba}+ f)}{(\omega_{ab}-\omega_{q})^{2}+ (\gamma_{ba}+ f)^{2}} \right )_{b=0, a=0}      }  \\ \nonumber
& & + \sum _{abc \neq 0} B_{q} \frac{(\gamma_{ba}+ f)}{(\omega_{ab}-\omega_{q})^{2}+ (\gamma_{ba}+ f)^{2}} \times  \\ \nonumber
& & \left ( 1+ C_{q} \frac{G_{q} (\gamma_{bc}+ f)}    {(\omega_{cb}-\omega_{p})^{2}+ (\gamma_{bc}+ f)^{2}} \right )
\end{eqnarray}

where $A_{q} = (N E_{q}\mu_{ab}^{i}\mu_{ba}^{j}\rho_{bb}^{(0)} )/ \hbar$ with $a,\ b,\ c=0$, $B_{q} = (N E_{q}\mu_{ab}^{i}\mu_{ba}^{j}(\rho_{aa}^{(0)}-\rho_{bb}^{(0)}) ) /\hbar $, $C_{q} = ( N E \mu_{cb}^{i}\mu_{bc}^{j}(\rho_{cc}^{(0)}-\rho_{bb}^{(0)}) )/\hbar $. At constant pump frequency the term containing $C_{q}$ is a constant , however it changes with pump field intensity. The first term in the right hand side of above equation is mainly responsible for the EIT transmission dip, depth of transmission increases linearly with $\rho_{bb}$ which in turn increases with pump field intensity.

When the laser with frequency $\omega_{q}$ is kept fixed in frequency and the laser $\omega_{p}$ is frequency scanned the absorption $\alpha(\omega_{p})$ can be expressed as follows


\begin{eqnarray}
\lefteqn {\alpha(\omega_{p})= \left ( -A_{p}\frac{(\gamma_{bc}+ f)}{(\omega_{cb}-\omega_{p})^{2}+ (\gamma_{bc}+ f)^{2}} \right )_{b=0, c=0}    }   \\ \nonumber
&+ \sum _{abc \neq 0} B_{p} \frac{(\gamma_{bc}+ f)}{(\omega_{cb}-\omega_{p})^{2}+ (\gamma_{bc}+ f)^{2}} \times \\ \nonumber
& \left ( 1+ C_{p} \frac{G_{p} (\gamma_{ba}+ f)}{(\omega_{ab}-\omega_{q})^{2}+ (\gamma_{ba}+ f)^{2}} \right )
\end{eqnarray}

where $A_{p} = (N E\mu_{cb}^{i}\mu_{bc}^{j}\rho_{cc}^{(0)} )/ \hbar$ with $a,\ b,\ c=0$, $B_{p} = (N E\mu_{cb}^{i}\mu_{bc}^{j}(\rho_{cc}^{(0)}-\rho_{bb}^{(0)}) ) /\hbar $, $C_{p} = ( N E \mu_{ab}^{i}\mu_{ba}^{j}(\rho_{aa}^{(0)}-\rho_{bb}^{(0)}) )/\hbar $ with $a,\ b,\ c \neq 0$. The expression for probe and coupling laser absorption for $^{85}Rb$ will look similar except the values of $a,\ b,\ c$.

\section{Experiment}

The experimental setup is schematically shown in figure 1b. Coupling laser (hereasfter called pump-1) is frequency locked to Rb transition line $^{87}Rb F=2  \rightarrow F^{\prime}=1 \ (^{85}Rb F=3 \rightarrow F^{\prime} = 2)$ through a separate saturation spectroscopy (figure 1b) while the probe laser (called probe-1) is frequency scanned over $\sim 500$ MHz width across $ F=1   \rightarrow F^{\prime}=x,\ (^{85}Rb F=3 \rightarrow F^{\prime} = x)\ x$ includes all excited state hyperfine levels. The pump and probe lasers are seperated using a polarizing cube beam splitter because they have orthogonal polarizations, the probe laser is fed onto  photodiode PD1 and EIT signal recorded.

Room temperature Rb cell is wrapped with a current carrying solenoid and placed inside a magnetic field shield so a uniform DC magnetic field is present and all random magnetic field eliminated. The magnetic field is to create the quantization axes along the beam propagation. A uniform axial magnetic field of about $400$ mG is produced by the solenoid. Pump laser optical intensity is controlled with the help of polarization cube beam splitter and half wave combination. Probe Rabi frequency kept fixed at $\Omega_{q}\sim 5 $MHz. Once the laser at $^{87}Rb F=2  \rightarrow F^{\prime}=1 $ transition acts as a coupling laser while probe laser scans across $^{87}Rb F=1   \rightarrow  F^{\prime}=0,\ 1,\ 2  $. In other set of measurement the laser at $^{87}Rb F=1  \rightarrow F^{\prime}=1 $ is frequency locked is called coupling laser 2 (or pump-2), whereas the other laser at $^{87}Rb F=2  \rightarrow F^{\prime}=1 $ is unlocked and its frequency is scanned across $^{87}Rb F=2  \rightarrow  F^{\prime}=1,\ 2,\ 3 $ and called probe-2.

In the same way pump-3 is locked to $^{85}Rb F=3 \rightarrow F^{\prime} = 2$ and pump-4 is frequency locked to $^{85}Rb  F=2 \rightarrow F^{\prime} = 2$. The probe-3,4 scanned across $F=2 \rightarrow F^{\prime}=x$ and $F=3 \rightarrow F^{\prime}=x$ respectively.

The pump and probe are mixed in copropagating configuration with the help of a cube beam splitter, so they remain orthogonally linearly polarized. At the exit from the EIT cell, another cube beam splitter seperates the pump and probe beams. The probe beam is incident on a photodiode and the photodiode response is recorded in a digital oscilloscope.

The laser locking is carried out in laser current modulation and analyzing the saturated absorption spectra in a lock-in-amplifier. Laser frequency is scanned with the help of piezo electric actuator.

\section{Results and Discussions}

As thermal vapor cell is used, having atomic number density of about $10^{10}$ per cm$^{3}$ with naturally abundant ratio of $^{85}Rb$ and $^{87}Rb$, so Doppler effect on EIT line width is obvious. Both pump and probe beam waist is 1 mm. Due to finite linewidth of the laser a group of atoms having certain velocity distribution will always be in resonance to the laser that will result in a broader EIT width. If the laser linewidth and  the atomic velocities were zero then a perfect quantum nature of the EIT width could be seen with ideal linewidth of the EIT spectra.

Transit relaxation originates from non-infinite stay time of atom to laser beam path and due to thermal speed of atom, this can be expressed as $\gamma_{t}= \frac{1}{2D}\sqrt{\frac{8k_{B}T}{\pi m}}$, where $D$ is optical beam diameter, $k_{B}$ Boltzmann constant, $T$ temperature of the cell, $m$ mass of Rb atom. For $2$mm beam diameter this becomes $15 \mu$s or $\gamma_{t}= 68.1$ KHz. Note that this is the maximum transit time related EIT width enhancement for atoms that move across the beam propagation, for other atoms the contribution will be lower as the atoms stay longer within the beam path.

It is to be noted that the equation (7) explains experiment for (pump-1, probe-1) and (pump-3, probe-3) configuration while equation (8) is for (pump-2, probe-2) and (pump-4, probe-4) or equation (7) is for figure 2, 4 while equation (8) is for figure 3, 5. Figure 2-5 shows the two EIT channels, each for $^{85}Rb$ and $^{87}Rb$ isotopes. Here we show that EIT can be observed either in the lower ground state or the upper ground state in $\Lambda$ system, depending upon which laser is frequency scanned. In all the cases the lasers were co-propagating and probe power Rabi frequency $\sim 5$MHz and detected by photodiode PD1. The corresponding probe saturated absorption spectra are shown in each graph as a dotted plot. The figures clearly demonstrate that EIT can occur in both the channels of $\Lambda$-type energy level configuration of $Rb$ atoms.To extrapolate this result further, we can predict that such a feature can be visible in any atom with $\Lambda$ configuration. Furthermore, considering the basic science behind the EIT phenomena, this observation also predicts that such a two channel EIT  is observable in any energy level configuration with any atomic vapor.

In figure 6a we show collection of EIT spectra for $^{87}Rb$ atoms with pump-1 and probe-1 configuration. The two broad absorption peaks correspond to absorption due to optical pumping of $Rb$ atoms to lower ground state HF levels and their separation is $157$ MHz, which is the frequency interval between $^{87}Rb, F^{\prime}=1\leftrightarrow F^{\prime}=2$. The EIT signal is observed at $ F=1  \rightarrow  F^{\prime}=1 $ transition, with a Lorentzian width of $\geq 300$ KHz.

 Figure 6(b) shows how Lorentzian width of transmission dip changes with $\Omega$, this is also predicted in equation (7).  The linear plot is the theoretical fit for  $f(\Omega)=0.386 + 3.0\times 10^{-2}\Omega$. It has also been known that such a linear change of EIT width take place with optical power [19], for example figure 3(b) of reference [17]. In addition to that our results in figure 6(c) shows change in depth of EIT as a function of $\Omega$ that also changes linearly, as predicted in equation (7).

  \begin{figure}[h]
  \includegraphics[angle=0,width=7cm]{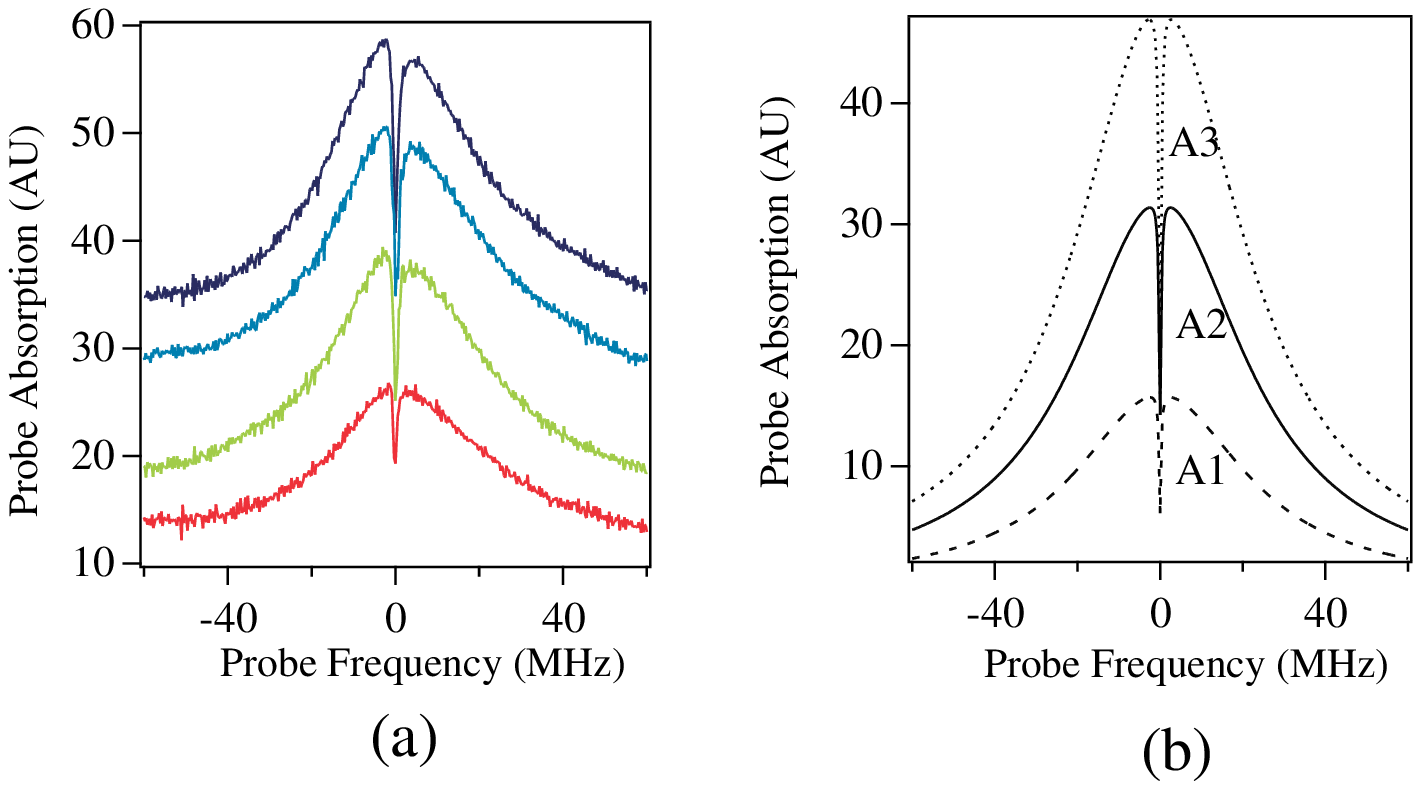}
  \caption{(a) Experimentally observed EIT signal as of figure 6a, (b) Theoretically calculated EIT signal at different pump laser power, $\Omega_{A3} =50,\ \Omega_{A2} = 25,\ \Omega_{A1}= 4$ MHz.}
 \label{fig5}
  \end{figure}

In equation (7), because of the factor $A(\gamma_{ba} + f)$ EIT peak depth is expected to change quadratically. From figure 6(c), the observed change in EIT depth $(\Delta f(\Omega))_{m_{F}=0} \sim 0.7 $ MHz for a change of pump Rabi frequency $\Delta \Omega \sim 25$MHz, or $\Delta \Omega \sim 30\times \Delta f$. Hence the effect of $f$ on non-linearity of EIT depth, as predicted in equation (7) is not very prominant, figure 6(c).

Figure 7(a) is experimental EIT spectra while 7(b) is drawn from theoretical calculation, using equation (7). To illustrate the properties of the spectra we set $\omega_{p}=\omega_{cb},\ A \sim 1,\ \gamma_{ba} +f = 330$KHz, for Zeeman sublevels equals to zero and for a combined contribution from all Zeeman sublevels not equals to zero $\gamma_{ba} +f = 25$ MHz. Keeping $ B\left( 1+ C \frac{G(\gamma_{bc}+ f)}{(\omega_{cb}-\omega_{p})^{2}+ (\gamma_{bc}+ f)^{2}} \right ) \sim 10^{4}$ as constant at a particular power, $\omega_{q}$ is varied, the resultant EIT spectra is as shown in figure 7(b). We assume $\rho_{aa},\ \rho_{cc},\ \rho_{bb}$ as unchanged. Note that excited state linewidth, as measured from absorption, is $\Gamma \sim 25 MHz$ whereas $\gamma_{0_{z}=0, m_{z}=0}\geq 300$ KHz.  

Similarly, it can be shown that a very similar nature of EIT spectra varying with coupling laser power for (pump-2, probe-2), (pump-3, probe3) and (pump-4, probe-4) systems.

\section{Conclusion}

We have shown experimentally and theoretically that EIT can be observed if coupling laser frequency is scanned and detected keeping its power low  as well as when the probe beam is scanned and detected. The good fitting between the experimental and theoretical simulation indicates Zeeman manifold of each hyperfine level has a significant role in EIT formation.

\section{Acknowledgements}
This work was supported by the Council of Scientific and Industrial Research of India.

\end{document}